\documentclass[pre,aps,amssymb,amsmath,superscriptaddress,twocolumn]{revtex4}

\usepackage{graphicx}
\usepackage{dcolumn}
\usepackage{bm}
\usepackage{epstopdf}

\newcount\driver
\newcount\bozza

scaled\magstep1                  
scaled\magstep1 \font\msytw=msbm10 scaled\magstep1
 
scaled\magstep1                 \font\indbf=cmbx10 scaled\magstep2

scaled \magstep2

{\count255=\time\divide\count255 by 60
\xdef\hourmin{\number\count255}
        \multiply\count255 by-60\advance\count255 by\time
   \xdef\hourmin{\hourmin:\ifnum\count255<10 0\fi\the\count255}}

\let\a=\alpha \let\b=\beta    \let\g=\gamma     \let\d=\delta     \let\e=\varepsilon
\let\z=\zeta  \let\h=\eta           \let\l=\lambda
\let\m=\mu    \let\n=\nu      \let\x=\xi                \let\r=\rho
\let\s=\sigma \let\t=\tau            \let\c=\chi
\let\ps=\psi   \let\o=\omega     
 \let\D=\Delta

\def\PP{{\cal P}}\def\EE{{\cal E}}\def\VV{{\cal V}}
\def\HH{{\cal H}}
\def\TT{{\cal T}}\def\NN{{\cal N}}
\def\RR{{\cal R}}\def\LL{{\cal L}}

\def\xx{{\bf x}}
  
\def\yy{{\bf y}}\def\nn{{\bf n}}
\def\zz{{\bf z}}
 \def\bP{{\bf P}}

       \def\oo{{\underline \omega}}
\def\ee{{\underline \varepsilon}}

\def\RRR{\hbox{\msytw R}}

        \def\EE{\hbox{\msytw E}}

\let\==\equiv

\let\io=\infty
\let\0=\noindent

\def\*{{\hfill\break\null\hfill\break}}

\def\tilde#1{{\widetilde #1}}

\def\aps{{\it a posteriori}}

\def\tende#1{\,\vtop{\ialign{##\crcr\rightarrowfill\crcr
             \noalign{\kern-1pt\nointerlineskip}
             \hskip3.pt${\scriptstyle #1}$\hskip3.pt\crcr}}\,}
\def\otto{\,{\kern-1.truept\leftarrow\kern-5.truept\to\kern-1.truept}\,}

\def\wh#1{\widehat{#1}}
\def\hat#1{\wh{#1}}
\def\sqt[#1]#2{\root #1\of {#2}}

\def\bp{{\bar \ps}}

\def\PP{{\cal P}}\def\EE{{\cal E}}\def\VV{{\cal V}}
\def\HH{{\cal H}}
\def\TT{{\cal T}}\def\NN{{\cal N}}
\def\RR{{\cal R}}\def\LL{{\cal L}}

\def\T#1{{#1_{\kern-3pt\lower7pt\hbox{$\widetilde{}$}}\kern3pt}}
\def\VVV#1{{\underline #1}_{\kern-3pt
\lower7pt\hbox{$\widetilde{}$}}\kern3pt\,}
\def\W#1{#1_{\kern-3pt\lower7.5pt\hbox{$\widetilde{}$}}\kern2pt\,}

\def\indica{\leaders \hbox to 0.5cm{\hss.\hss}\hfill}
\def\guida{\leaders\hbox to 1em{\hss.\hss}\hfill}
\mathchardef\oo= "0521

\def\xx{{\bf x}}
\def\yy{{\bf y}}\def\nn{{\bf n}}
\def\zz{{\bf z}}
 \def\bP{{\bf P}}
 
\def\oo{{\underline \omega}}

\def\qed{\raise1pt\hbox{\vrule height5pt width5pt depth0pt}}
 
  \def\bp{{\bar p}} 

\def\indic{\hbox{\raise-2pt \hbox{\indbf 1}}}

\def\RRR{\hbox{\msytw R}}

\def\ins#1#2#3{\vbox to0pt{\kern-#2 \hbox{\kern#1 #3}\vss}\nointerlineskip}

\newdimen\xshift \newdimen\xwidth \newdimen\yshift

\def\insertplot#1#2#3#4#5#6{%
\xwidth=#1pt \xshift=\hsize \advance\xshift by-\xwidth \divide\xshift by 2%
\begin{figure}[ht]
\vspace{#2pt} \hspace{\xshift}
\begin{minipage}{#1pt}
#3 \ifnum\driver=1 \griglia=#6
\ifnum\griglia=1 \openout13=griglia.ps \write13{gsave .2
setlinewidth} \write13{0 10 #1 {dup 0 moveto #2 lineto } for}
\write13{0 10 #2 {dup 0 exch moveto #1 exch lineto } for}
\write13{stroke} \write13{.5 setlinewidth} \write13{0 50 #1 {dup 0
moveto #2 lineto } for} \write13{0 50 #2 {dup 0 exch moveto #1
exch lineto } for} \write13{stroke grestore} \closeout13
\includegraphics{griglia.ps} \fi
\includegraphics{#4.ps}\fi%
\ifnum\driver=2 \fi
\end{minipage}
\caption{#5}
\end{figure}
}

\newdimen\shift \shift=-1.5truecm
\def\lb#1{%
\ifnum\bozza=1
\label{#1}\rlap{\hbox{\hskip\shift$\scriptstyle#1$}}
\else\label{#1} \fi}

\def\be{\begin{equation}}
\def\ee{\end{equation}}
\def\bea{\begin{eqnarray}}\def\eea{\end{eqnarray}}
\def\bean{\begin{eqnarray*}}\def\eean{\end{eqnarray*}}
\def\bfr{\begin{flushright}}\def\efr{\end{flushright}}
\def\bc{\begin{center}}\def\ec{\end{center}}
\def\bal{\begin{align}}\def\eal{\end{align}}
\def\ba#1{\begin{array}{#1}} \def\ea{\end{array}}
\def\bd{\begin{description}}\def\ed{\end{description}}
\def\bv{\begin{verbatim}}\def\ev{\end{verbatim}}
\def\nn{\nonumber}
\def\Halmos{\hfill\vrule height10pt width4pt depth2pt \par\hbox to \hsize{}}
\def\pref#1{(\ref{#1})}

\def\ins#1#2#3{\vbox to0pt{\kern-#2 \hbox{\kern#1 #3}\vss}\nointerlineskip}

\newdimen\xshift \newdimen\xwidth \newdimen\yshift
\newcount\griglia

\def\insertplot#1#2#3#4#5#6{%
\xwidth=#1pt \xshift=\hsize \advance\xshift by-\xwidth \divide\xshift by 2%
\begin{figure}[ht]
\vspace{#2pt} \hspace{\xshift}
\begin{minipage}{#1pt}
#3 \ifnum\driver=1 \griglia=#6
\ifnum\griglia=1 \openout13=griglia.ps \write13{gsave .2
setlinewidth} \write13{0 10 #1 {dup 0 moveto #2 lineto } for}
\write13{0 10 #2 {dup 0 exch moveto #1 exch lineto } for}
\write13{stroke} \write13{.5 setlinewidth} \write13{0 50 #1 {dup 0
moveto #2 lineto } for} \write13{0 50 #2 {dup 0 exch moveto #1
exch lineto } for} \write13{stroke grestore} \closeout13
\includegraphics{griglia.ps} \fi
\includegraphics{#4.ps}\fi%
\ifnum\driver=2 \fi
\end{minipage}
\caption{#5}
\end{figure}
}

\newdimen\shift \shift=-1.5truecm
\def\lb#1{%
\label{#1}\rlap{\hbox{\hskip\shift$\scriptstyle#1$}}
\else\label{#1} \fi}

\def\be{\begin{equation}}
\def\ee{\end{equation}}
\def\bea{\begin{eqnarray}}\def\eea{\end{eqnarray}}
\def\bean{\begin{eqnarray*}}\def\eean{\end{eqnarray*}}
\def\bfr{\begin{flushright}}\def\efr{\end{flushright}}
\def\bc{\begin{center}}\def\ec{\end{center}}
\def\bal{\begin{align}}\def\eal{\end{align}}
\def\ba#1{\begin{array}{#1}} \def\ea{\end{array}}
\def\bd{\begin{description}}\def\ed{\end{description}}
\def\bv{\begin{verbatim}}\def\ev{\end{verbatim}}
\def\nn{\nonumber}
\def\Halmos{\hfill\vrule height10pt width4pt depth2pt \par\hbox to \hsize{}}
\def\pref#1{(\ref{#1})}

\usepackage{amsmath}
\usepackage{amsfonts}
\usepackage{amssymb}
\usepackage{epstopdf}

\font\msytw=msbm9 scaled\magstep1 
scaled\magstep1

\let\a=\alpha \let\b=\beta  \let\g=\gamma  \let\d=\delta
\let\e=\varepsilon
\let\z=\zeta  \let\h=\eta     \let\l=\lambda
\let\m=\mu    \let\n=\nu    \let\x=\xi         \let\r=\rho
\let\s=\sigma \let\t=\tau    \let\c=\chi
\let\ps=\Psi   \let\o=\omega
 \let\D=\Delta

\def\EE{{\cal E}} \def\VV{{\cal V}}
 
\def\TT{{\cal T}}\def\NN{{\cal N}} 
\def\RR{{\cal R}}\def\LL{{\cal L}}

 \def\xx{{\bf x}} \def\yy{{\bf y}} \def\zz{{\bf z}}

\def\PP{{\bf P}}

\def\nn{\nonumber}

\def\RRR{\hbox{\msytw R}}

\def\\{\hfill\break}
\def\={:=}
\let\io=\infty
\let\0=\noindent

\def\tende#1{\,\vtop{\ialign{##\crcr\rightarrowfill\crcr\noalign{\kern-1pt
    \nointerlineskip} \hskip3.pt${\scriptstyle #1}$\hskip3.pt\crcr}}\,}
\def\otto{\,{\kern-1.truept\leftarrow\kern-5.truept\to\kern-1.truept}\,}

\def\wh{\widehat}
\def\to{\rightarrow}

\def\qed{\hfill\raise1pt\hbox{\vrule height5pt width5pt depth0pt}}

\def\be{\begin{equation}}
\def\ee{\end{equation}}
\def\bp{\begin{pmatrix}}
\def\ep{\end{pmatrix}}
\def\bea{\begin{eqnarray}}
\def\eea{\end{eqnarray}}
\def\nn{\nonumber}
\def\pref#1{(\ref{#1})}

\def\lb{\label}

\newtheorem{lemma}{Lemma}[section]

\begin{document}

\title{Localization of interacting fermions in the Aubry-Andr\'e model}
\author{Vieri Mastropietro}
\affiliation{Universit\`a degli Studi di Milano, 
Via Saldini, 50, 20133 Milano - Italy}

\begin{abstract}  
We consider interacting electrons in a one dimensional lattice
with an incommensurate Aubry-Andr\'e potential
in the regime when the single-particle eigenstates are localized. We rigorously establish
persistence of ground state localization in presence of weak many-body interaction, for 
almost all the chemical potentials. The proof uses a quantum many body extension of methods adopted for the stability of tori of nearly integrable hamiltonian systems, 
and relies on number-theoretic properties of the potential incommensurate frequency. 
\end{abstract}


\maketitle


Interacting fermions with a quasi-periodic potential describe quasicrystals or
crystals with incommensurate charge density waves, and have a number of features in common
with systems with random disorder; in particular  {\it Anderson localization} \cite{A} can occur in
the single particle eigenstates. The 
paradigmatic model for such systems
is the  interacting
{\it Aubry-Andr\'e}  model \cite{AA}, which can be equivalently written as the Heisenberg XXZ spin chain
with an incommensurate  magnetic field. The interest of this model relies from one side on the fact that, despite its simplicity,
it has a number of non trivial features common to more realistic systems, like a metal-insulator transition even at the single particle level; 
on the other, it provides an accurate description of  
cold atoms with quasi-random optical lattices
generated by two or more incommensurate frequencies
\cite{I1},\cite{I3},\cite{I2}. At the single particle level, that is neglecting 
the many body interaction, its properties follow from 
the {\it Harper} or {\it almost-Mathieu} equation which have been extensively studied
in the last thirty years; it is known
in particular that the
spectrum is a Cantor set \cite{A1} with a dense set of gaps 
and that its eigenfunctions are extended \cite{Si} or localized 
\cite{FSW}  varying the strength of the potential. 
Such remarkable properties are
related to a deep connection 
between the non interacting  Aubry-Andr\'e model and the {\it Kolnogorov-Arnold-Moser} (KAM) theorem
expressing the stability of invariant tori in quasi integrable classical hamiltonian systems.

Much less is known when a many-body interaction is present, and one of 
the key question is
which is the fate of Anderson localization, which is a single particle phenomenon. Such problem is of general interest
in disordered quantum matter (a quasi-periodic potential can be considered a quasi-random disorder);
in the case of
truly {\it random} disorder
analytical evidence supporting localization in presence of interaction
has been obtained \cite{GGG}, \cite{B}, \cite{loc4}, 
and consequences for dynamics has been exploited \cite{PH1},\cite{PH2},\cite{PH4},\cite{PH5},\cite{NH}; 
a complete
proof of localization in presence of interaction, even if limited to the ground state, is 
however still lacking.
In the interacting  Aubry-Andr\'e model, analytical results have been established
\cite{M}, \cite{G2} only in the {\it extended} regime; in the {\it localized} phase
numerical simulations 
\cite{H4} and cold atoms experiments \cite{I2} support the stability of localization
for attractive or repulsive interactions.

In this letter we rigorously establish in the Aubry-Andr\'e model the
persistence of ground state localization in presence of weak many-body interaction, for 
almost all the chemical potentials. 
The analysis relies heavily on number-theoretical properties of the frequency of the incommensurate potential, which play a key role
already in the non interacting case; however the structure of small divisors is made involved by the presence of loops
caused by the many body interaction. Our result is obtained by a new technique based on a combination 
of multiscale methods developed for the classical Lindstedt series for KAM invariant tori in nearly integrable Hamiltonian systems \cite{G} combined with non-perturbative Renormalization Group (RG) methods \cite{GM}.

The Hamiltonian of the interacting Aubry-Andr\'e model is
\bea &&H=-\e(\sum_x (a^+_{x+ 1} a_{x}+ a^+_{x} a^-_{x-1} )+\m \sum_x 
 a^+_{x} a^-_{x}
\nonumber\\
&&+u\sum_x  \phi_x
a^+_{x} a^-_{x}+U \sum_{x}
a^+_{x} a^-_x a^-_{x+1} a^+_{x+1}
\label{1.1}\eea
with $a^\pm_x$ fermionic creation or annihilation operators, $x$ points on a one dimensions lattice  
with step $1$ and Dirichelet boundary conditions
and $\phi_x$ is a potential {\it incommensurate} with the lattice
$\phi_x=\cos (2\pi \o x)$ and $\o$ {\it irrational}; it is convenient, as in the analysis 
of the Harper
equation, to assume that
$\o$ verifies a {\it Diophantine} condition (valid on a full measure set)
\be ||2\pi n\o||\ge C
n^{-\t}\label{d}\ee for any integer non vanishing $n$ ($||.||$ is the norm on the one dimensional torus).
For definiteness we are choosing coordinates so that $\phi_x$ is even with respect to $x=0$. 

Ground state localization implies zero temperature
exponential decay for large values of the coordinate difference in the thermodynamical correlations,
for instance in  the 2-point function 
$<T a^-_\xx a^+_\yy>$, where $T$ is the time order product, $a_\xx^\pm=e^{H x_0} a^\pm_x e^{-H x_0}$,
$\xx=(x_0,x)$
and 
$<>={{\rm Tr} e^{-\b H} .\over{ \rm Tr} e^{-\b H}}$. In the absence of hopping or interaction $U=\e=0$
({\it molecular limit})
there is perfect localization and, if $<T a^-_\xx a^+_\yy>|_{U=\e=0}=g(\xx,\yy)$
\be
g(\xx,\yy)=\d_{x,y}\int dk_0 
{e^{ik_0(x_0-y_0)} 
\over -i k_0+u\phi_x -\m}
\label{prop}
\ee
which is
is exactly vanishing for fermions with 
different coordinates. 
We can now consider small 
$U,\e$ (compared to $\e$) and see if the theory is 
{\it analytically close} to the molecular limit.
The interacting 2-point function can be written in terms of a perturbative expansion
in $U,\e$ and
each term of the series
is expressed in terms of Feynman graphs given by sums of products of propagators \pref{prop}, 
whose singularities 
naturally determine the physical properties.

As the frequency $\o$ is irrational, $(\o x)_{mod. 1}$
fills densely the interval $(-1/2,1/2]$ so that the denominator $u\phi_x -\m$
can be {\it arbitrarily small}; this happens when 
 $(\o x)_{mod. 1}$ is close to $\pm (\o \bar x)_{mod. 1}$ with
$\m=u \cos 2\pi \o \bar x$; then  for small $(\o x')_{\rm mod. 1}$ one has 
$\phi_{x'\pm \bar x}-\m\sim \pm  (\o x')_{\rm mod. 1}$.
Note the close analogy between the 2-point function \pref{prop} at the {\it molecular limit} $u=\e=0$,  with the propagator in the 
{\it free Fermi gas} limit  $U=u=0$, namely  $(-i k_0+\e\cos k-\m)^{-1}$; the points $\pm \bar x$
have the same role as the Fermi momenta $\pm p_F$ such that $\m=\cos p_F$, so it is natural to call them {\it Fermi coordinates}.
This relation between molecular and free gas limit is a manifestation of a property of the Harper equation
known as {\it Aubry duality}.

We expect that  the interaction causes a renormalization of the chemical potential
and
it is then convenient to {\it fix} the interacting chemical potential by choosing properly its bare value; in particular the bare chemical potential is taken as 
$\m=u \cos (2\pi \o\bar x)+\e\n$, and $\n$ is determined by requiring the the dressed chemical potential is $u\cos (2\pi \o \bar x)$. There are two main cases to be considered, corresponding to a choice of the chemical potential in a point of the non interacting spectrum
or in a gap (which are a dense set).
The first case is when $\bar x$ is irrational; in this case 
fermions living close to different Fermi points, that is with 
coordinate $x'-\bar x$ and $x'+\bar x$ with $(\o x')_{mod. 1}$ small, cannot be {\it exactly} connected by the hopping or
by the interaction, which causes jumps of $0,\pm 1$ sites (but they can be arbitrarily close connected). 
Again it is convenient to require a strong irrationality condition \cite{Si} valid on a {\it full measure} set, namely
\be ||\o x\pm 2 \o \bar x||\ge C |x|^{-\t}\label{fg}\ee
The second case is when opposite Fermi coordinates can be exactly connected; this happens
when $2\bar x$ is an integer and causes the gaps in the non interacting spectrum, exactly like a periodic potential produces gaps in the free Fermi gas limit (in space of momenta instead of coordinate).
A consequence of the above choices of $\bar x$ is that
$
|\hat g(k_0, x)|\le C |x|^\t$, 
that is the denominator can be very small for large $x$; this is a sort of {\it ultraviolet-infrared} mixing, a
phenomenon typical of incommensurate potentials. 
Our main result is the following.
\vskip.3cm
{\rm Theorem.} {\it Let us fix $u=1, U=\l\e$, and assume $\e,\l$ small
and $\o$ verifying \pref{d}, $\m=\phi_{\bar x}+\e \n$. Then if
$\bar x$ verifies \pref{fg} (chemical potentials outside the gaps of the non interacting spectrum)
or f $\bar x$ is half integer (chemical potentials in the gaps of the non interacting spectrum), then, for a suitable choice of $\n$, 
the zero temperature 2-point function obeys to
\be|<T a^-_\xx a^+_\yy>|\le C e^{- \x|x-y|} F(x_0-y_0)\label{fon}
\ee
with $\x=C |\log\e|$, $C$ a constant, and $F= {1\over 1+(\D|x_0-y_0)|)^N}$ 
with $\D=(1+min(|x|,|y|))^{-\t}$
if $\bar x$ verifies \pref{fg}, while 
$\D=a \e^{2\bar x}+O(\e^{2\bar x+1})$ and $a\not=0$ if $\bar x$ is half integer.
}
\vskip.3cm
The above result establishes exponential decay in the 2-point function,
implying Anderson localization for the ground states
of the interacting system 
with chemical potentials in the center of the gaps or in points of the spectrum verifying
\pref{fg}; in the first case
there is a fast decay in time with rate proportional to the gap size. 
\vskip.2cm
{\it Proof.} We present here the key steps of the proof (more technical details will be published elsewhere \cite{M}).
The 2-point function is obtained by the second derivative of the generating function 
\be
e^{W(\phi)}=\int P(d\psi)e^{V(\psi)+(\psi,\phi)}\label{ww}
\ee
with
\bea
&&V=-\l\int d\xx \psi^+_{\xx}\psi^-_{\xx} \psi^+_{\xx+{\bf e}_1}\psi^-_{\xx+{\bf e}_1}+\\
&&\e\int d\xx (
\psi^+_{\xx} \psi^-_{\xx+{\bf e_1}}+\psi^+_{\xx+{\bf e_1}}\psi^+_{\xx})+\n \int d\xx
\psi^+_{\xx} \psi^-_{\xx}\nonumber
\eea
where $\psi$ are grassmann variables, $\phi$ is the external source, $\int d\xx=\int dx_0\sum_x$,
${\bf e}_1=(0,1)$ and $P(d\psi)$ is the fermionic integration with propagator 
\pref{prop}. We introduce
a cut-off smooth function $\chi_\r(k_0,x)$ which is non vanishing for
$\sqrt{k_0^2+((\o(x-\r\bar x)_{\rm mod. 1})^2)}\le \g$, where $\r=\pm 1$ and $\g>1$
is a suitable constant (to be fixed below); therefore we can write the propagator as 
\be
\hat g(k_0,x)=\hat g^{(u.v.)}(k_0,x)+\sum_{\r=\pm}\hat g_\r(k_0, x)
\ee where $\hat g_\r(k_0, x)={\chi_\r(k_0,x)\over -i k_0+\phi_x-\m}$, 
and correspondingly
 $\psi_{k_0,x}=\psi^{(u.v.)}_{k_0, x}+\sum_{\r=\pm 1} \psi_{\r, k_0, x}$. This simply says that we are rewriting 
the fermionic field as sum of two independent fields living close to one of the Fermi points, up to a regular field.
We can further decompose \be \hat g_\r(k_0, x)=\sum_{h=-\io}^0 \hat g^{(h)}_{\r}( k_0,x)\ee with 
$\hat g^{(h)}_\r(k_0,x)$ similar to $\hat g^{(h)}_{\r}(k_0,x)$ with $\chi$ replaced by $f_h$ 
with
where $f_h(k_0,\0 x')$ is non vanishing in a region $\sqrt{k_0^2+((\o x')_{\rm mod 1})^2}\sim \g^h$. 
After the integration of $\psi^{(u.v.)}, \psi^{(0)},..,\psi^{(h+1)}$ the generating function has the form
\be
e^{W(\phi)}=\int P(d\psi^{\le h})e^{V^{(h)}(\psi)+B^{(h)}(\psi,\phi)}\label{hh}
\ee
where $P(d\psi^{\le h})$ has propagator $g_\r^{(\le h)}=\sum_{k=-\io}^h g_\r^{(k)}$ and $V^{(h)}(\psi)$
is given by
\be
V^{(h)}(\psi)=\sum_m W_m^{(h)}
\psi^{\e_1(\le h)}_{\r_1,x_{0,1},x'_1+\r_1\bar x}... \psi^{\e_m(\le h)}_{\r_m,x_{0,m},x'_m+\r_m\bar x}
\label{ep}\nn\ee
and the kernels $W_m^{(h}$ are sum of Feynman diagrams obtained connecting vertices $\e$, $\l$
or $\n$ with propagators $g^{(k)}$ with $k>h$ ; $B^{(h)}$ is given by a similar expression with the only difference that some of the external lines are associated to $\phi$ fields.
In each of the Feynman diagrams contributing to $W_m^{(h)}$ there is a
 {\it tree} of propagators connecting all the external lines and the vertices; the coordinates $x_i$, $x_j$ of two external lines
as such that 
\be x_i-x_j=x'_i+\r_i\bar \o-x'_j-\r_j\bar \o=
\sum^*_\a \d_\a\label{hhh}\ee 
where the sum is over the vertices in the path of the tree connecting $i$ and $j$ and $\d_\a=(0,1,-1)$ (see Fig. 1) is associated to the line connected to the vertex $\a$.  Of course $x_i-x_j$ has an integer value. 
\vskip0.9cm
\insertplot{220}{125}
{\ins{100pt}{110pt}{$w_1$}
\ins{75pt}{100pt}{$w_a$}
\ins{45pt}{80pt}{$w_b$}
\ins{20pt}{80pt}{$w_c$}
\ins{40pt}{40pt}{$w_2$}
}
{fig60}
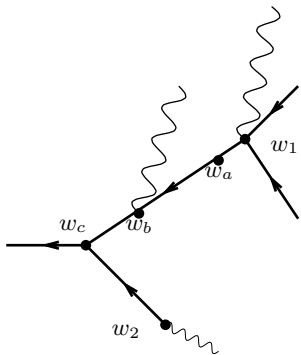
{An example of a tree; the solid lines represent propagators and the wiggly lines the external lines;  $x_{w_1}-x_{w_2}
=\d_1+\d_a+\d_b+\d_c+\d_2$.
} {0}
Each kernel $W^{(h)}_m$ is expressed by a power series in $\e,\n,\l$, and one has to show convergence uniformly in
the RG step $|h|$; however 
standard {\it power counting} arguments says that the theory is {\it non renormalizable}, as the naive degree of divergence is $k-1$ if  $k$ is the order of the graph. Power counting in non sufficient
and one has to use a more accurate analysis
exploiting the number-theoretic properties of the frequency.

In order to do that it is convenient, 
given a Feynman graph, 
to consider a maximally connected subset of lines corresponding 
to propagators 
with scale $h\ge h_v$ with at least a scale $h_v$, and we call it {\it cluster} $v$; 
the external lines have scale smaller then $h_v$. Therefore to each Feynman graph is associated a hierarchy of 
clusters; inside each cluster $v$ there are $S_v$ maximal clusters, that is clusters   
contained only in the cluster $v$ and not in any smaller one, or trivial clusters given by a single vertex. Each of such $S_v$ clusters are connected 
by a tree of propagators with scale $h_v$; by integrating the propagators over the time, and using that 
$\int dx_0 |g^{(h)}_\r(x_0, x)|\le C\g^{-h}$ and that $|g^{(h)}_\r(x_0, x)|\le C$ we get 
that each graph of order $n$ contributing to $W^{(h)}_m$ is bounded by \be C^n \e^n
\prod_v \g^{-h_v(S_v-1)}\label{paz}\ee
where $v$ are the clusters and
$h_v\le 0$. As $S_v\ge 1$ the above estimate says that the size of the graph increases with $|h|$
and
diverges in the zero temperature limit.
In getting \pref{paz}, 
we have however not used a crucial property implied by the Diophantine condition: namely
that if the denominators associated to the external lines have the same size, the difference of coordinates must be or 
$0,\pm 2\bar x$ (the second case only for $\bar x$ half integer)
 or very large; in this second case there is a large number of hopping or interactions terms by \pref{hhh} and therefore a decaying factor associated to an high power of $\e$.
This suggests to distinguish in $V^{(h)}$
two kinds of terms, the {\it resonant terms}, such that the coordinate $x'_i$ measured from the Fermi coordinates of the external fields are all {\it equal}, and the remaining {\it non resonant} terms.

We define, as usual in the theory of renormalization, a localization operator 
$\LL$ acting only on the resonant terms and setting all the temporal coordinates of the external fields equal.
%
We split the resonant terms in a {\it local} part, where $\LL$ applies, 
and a {\it renormalized} part, where
$\RR=1-\LL$ applies.
In the $\RR$ part  there is at least a difference of fields 
$(\psi^{\e(\le h)}_{\r,x_{0,i},x'+\r\bar x}-\psi^{\e(\le h)}_{\r,x_{0,i},x'+\r\bar x})$
and this produces a extra gain $\g^{h_{v'}-h_v}$, if $v$ is a resonant cluster and $v'$ 
is the minimal cluster containing it. Regarding the local terms, they are proportional to monomial of fields with the same coordinates $\xx'$, hence anticommutativity implies that terms with more than four fields are exactly
vanishing. The possibly non vanishing quartic terms have the form
\be \e\l_h \int d\xx \psi^+_{+,x_0,x'+\bar x} \psi^-_{+,x_0,x'+\bar x}\psi^+_{-,x_0,x'-\bar x} 
\psi^-_{-,x_0,x'-\bar x}\ee
When $\bar x$ is irrational such terms are vanishing as there are fields with coordinate difference $2\bar x$ 
which by \pref{hhh} should be an integer. When $\bar x$ is half-integer local quartic terms exist 
and in such a case $\l_h=O(\e^{2\bar x-1}\l)$.
Similarly the quadratic terms are of two kinds; one is
$\e
\n_h\sum_\r \int d\xx \psi^+_{+,x_0,x'+\r\bar x} \psi^-_{+,x_0,x'+\r\bar x}$
representing the renormalization of the chemical potential, the 
other is
$\s_h \sum_\r\int d\xx \psi^+_{+,x_0,x'+\r\bar x} \psi^-_{+,x_0,x'-\r\bar x}$
and is vanishing when $\bar x$ is irrational, while is present
when $\bar x$ is half integer and in such case $\s_h=O(\e^{2\bar x})$.

We have now to consider the contributions from the {\it non resonant} terms of the effective potential $V^{(h)}$.
In such terms there are least two external fields with coordinate $x'_1, x'_2$ with $x'_1\not= x'_2$; 
if $|h|$ is large the corresponding divisors are small $O(\g^h)$ but the coordinate difference is large. More quantitatively, if $m=\sum^*_\a \d_\a\not=0$, see \pref{hhh} and $v'$ is the cluster containing $v$
\bea
&&2\g^{h_{v'}}\ge ||(\o x'_1)||+||(\o x'_2)||\ge ||
\o(x'_1-x'_2)||= \nn\\
&&||\o(\r_2-\r_1)\bar x+m\o ||\ge C_0 |\d 2\bar x+|m||^{-\t}
\label{ll}\eea
with $\d=0$ when $\bar x$ is irrational and $\d=1$  when is half integer; the last inequality follows from
\pref{d} and \pref{fg}. Therefore $|m|\ge \tilde C \g^{-h/\t}$, that is $m$ must be very large
if the divisors are small and not coinciding. In a non resonant cluster there then are necessarily 
a large number of vertices $N$ so that
$N\ge |m|\ge \tilde C \g^{-h/\t}$; by taking into account the hierarchy 
of
clusters one gets \be \e^n\le \e^{n\over 2}\prod_v \e^{C 2^{h_v}\g^{-h_v/\t}
S^{NR}_v}\ee
where $S^{NR}_v$ are the non resonant clusters contained in $v$;
that is a decay factor is associated to each non resonant cluster.
The power counting bound \pref{paz}, apparently saying that the theory is non renormalizable, is therefore replaced by
\be
G
\e^{n\over 2}
[\prod_v \g^{-h_v(S_v-1)} ]
[\prod_v \e^{C 2^{h_v}\g^{-h_v/\t}S^{NR}_v}] [\prod^*_{v}\g^{h_{v'}-h_v}] \label{ggh}\ee
where $\prod^*_{v}$ is over the resonant terms
and $G$ contains the product of running coupling constants. 
Note that $S_v=S^{R}_v+S^{NR}_v+S^e_v$, where $S^R_v$ is the number of resonant clusters and 
$S^e_v$ is the number of end-points with scale $h_v$ in the cluster $v$. 
The above bound is true not only for the single graph, but for the sum of all graphs contributing to 
order $n$ and summing over all the possible choices of external lines of the clusters.
Note that $[\prod_v \g^{-h_v(S_v^R-1)} ][\prod^*_{v}\g^{h_{v'}-h_v}]\le 1$.
Finally one has to discuss the flow of the effective running coupling constant. When $\bar x$ is irrational the only running coupling constant is $\n_h$, expressing the renormalization of the chemical potential
and by choosing properly $\n$ then by a fixed point argument $\n_h=O(\g^h)$. When $\bar x$ is half integer
then one has also a mass term $\s_h=a \e^{2\bar x}+(\e^{2\bar x+1})$ and this implies that there is a mass scale $h^*$ of the order of the log of the mass. The effective coupling is $\l_h=\l_0(1+O(|h^*|\e)$
and as $\l_0=O(\e^{2\bar x-1}\l)$ then $\g^{-h}|\e\l_h|$ is small. From 
\pref{ggh} we have finally that the size of the $n$-th order contibution to $W_m^{(h)}$ is
\be\e^{n\over 2}
[\prod_v (\g^{-h_v}\e^{C 2^{h_v}\g^{-h_v/\t}})^{S^{NR}_v}]\le C^n \e^{n\over 2}
[\prod_v \g^{h_v S^{NR}_v}]\ee
If $\g^{1\over \t}>2$ and $\e$ small one
can then sum over $\{h_v\}$ and get an $\e^{n\over 2} C^n$ bound. When $S^{NR}_v\not=0 $
this is trivial. If $S^{NR}_v=0$ then $v$ is resonant and:
a)
when $\bar x$ is rational one uses the gap saying that the number of scales is $O(\log\e)$; b) when $\bar x$ is irrational,  if $v$ has two external lines than there is no sum as $h_{v}=h_{v'}+1$ 
by the support properties of $f_h$ 
as the internal lines of $v$ has the same $k_0, x'$ as the external ones; if has more than four lines, $v$ is resonant and its local part is vanishing so one has an extra $\g^{h_{v'}-h_v}$ allowing the sum over $h_v$.
A similar analysis can be done for the terms with 2 external lines $\phi$ contributing to the 2-point function; the exponential decay in the coordinate follows again from \pref{hhh} as the only non vanishing contributions have an order greater than $|x-y|/2$.

In conclusion we have rigorously established 
persistence of localization of the ground states for almost all the chemical potentials
and in presence of attractive or repulsive weak interaction. The analysis 
relies heavily on the number theoretic properties of the frequency of the quasi-periodic potential. 
Our result would imply 
localization of all the eigenfunctions 
in a quasi-free theory, and we believe that an extension of the methods 
introduced here will
be able to establish localization also at finite temperature, or extended to random disorder.

\bibliographystyle{amsalpha}

\end{document}